\documentclass[aps,pra,reprint,groupedaddress,twocolumn]{revtex4-2}
\usepackage{graphicx}
\usepackage{xcolor}
\usepackage{amsmath}
\usepackage{mathtools}
\usepackage{amssymb}
\DeclareMathOperator*{\argmax}{arg\,max}

\begin{document}

\title{Quantum Target Ranging for LiDAR }
\author{Giuseppe Ortolano$^{1,2,3}$}
\author{Ivano Ruo-Berchera$^1$}
\affiliation{$^1$Quantum metrology and nano technologies division, INRiM, Strada delle Cacce 91, 10153 Torino, Italy}
\affiliation{$^2$ Dipartimento di Fisica e Astronomia, Università di Firenze, Via G. Sansone 1, I-50019 Sesto Fiorentino (FI), Italy}
\affiliation{$^3$ Istituto Nazionale di Fisica Nucleare, Sezione di Firenze, via G. Sansone 1, I-50019 Sesto Fiorentino (FI), Italy}

\begin{abstract}
We investigate Quantum Target Ranging in the context of multi-hypothesis testing and its applicability to real-world LiDAR systems. First, we demonstrate that ranging is generally an easier task compared to the well-studied problem of target detection. We then analyze the theoretical bounds and advantages of quantum ranging in the context of phase-insensitive measurements, which is the operational mode of most LiDAR systems. Additionally, we adopt a background noise model more suited to optical frequencies, as opposed to the typical single-mode thermal noise model used in quantum target detection theory. Our findings indicate that a significant exponential quantum advantage can be achieved using simple photon-counting receivers across a broad range of parameters, thereby validating the efficacy of the quantum approach for LiDAR implementations.
\end{abstract}

\maketitle

In the context of quantum sensing \cite{Degen_2017,Pirandola_2018,Petrini_2020,Banchi_2020,Barbieri_2022,Berchera_2013,Berchera_2019,Ortolano_2023a,Pereira_2023,Ortolano_2023b}, improving target detection and ranging by leveraging quantum correlations and entanglement has received large attention in the last 15 years \cite{Maccone_2020, Barzanjeh_2015, Sorelli_2022,Zhao_2022,Torrome_2024}, stemming from the proposal of the quantum illumination (QI) protocol \cite{Sacchi_2005,Lloyd_2008}. 
The huge interests is justify by the perspective of a potential disruptive advancement of the radar and LiDAR technology \cite{Schwarz_2010, Hong_2024}, widely used in surveillance, both civil and military aviation, automotive, environment monitoring and even biological imaging.

The problems of quantum target detection (QTD)  and ranging (QTR) are closely related. In QTD one uses a probe state to asses whether a target is present or not at a given \emph{a priori} position. On the other hand, in QTR one assumes that the object is present and tries to localize it.  It is clear that any real world application would most likely need to address both tasks. 

\begin{figure*}[th]
	\includegraphics[height=0.3\textwidth,width=\textwidth]{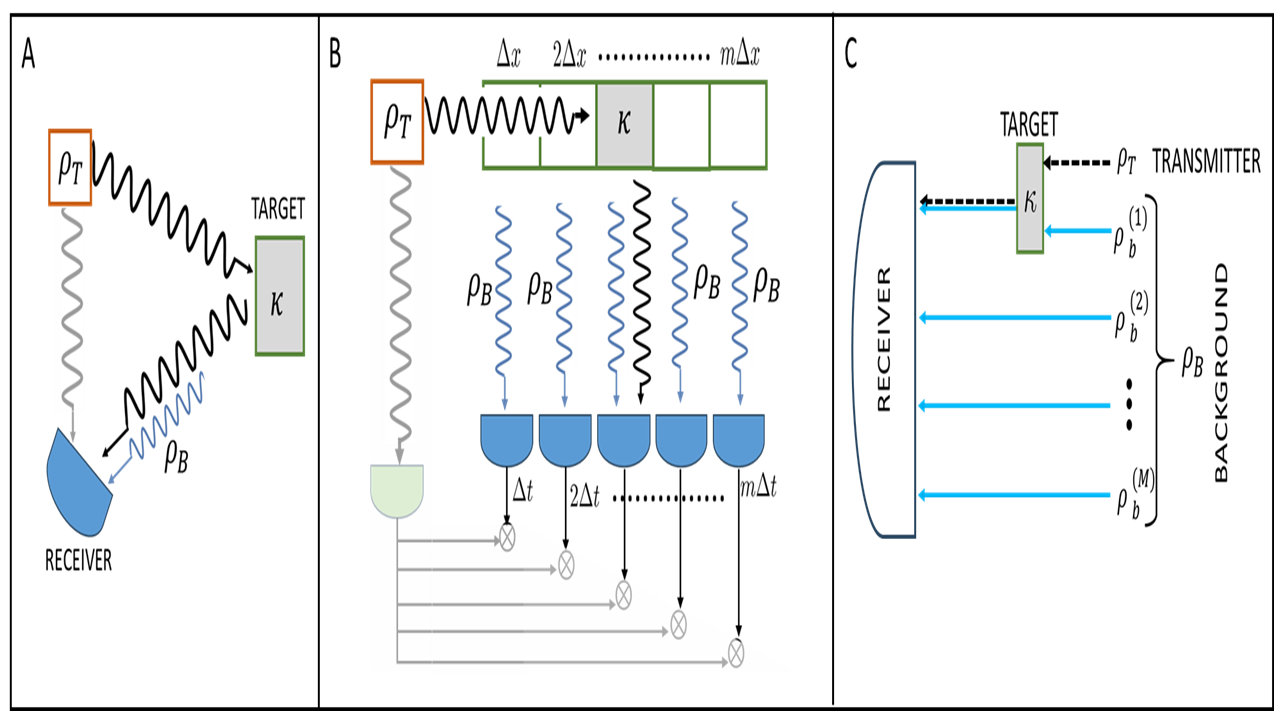}
	\caption{\textit{Sensing and detention schemes.} \textbf{A.} QTD: A transmitter signal mode in the state $\rho_T$ interacts with a target of reflectance $\kappa$ and reflected back at the receiver immersed in a background. An ancillary mode (in faint color) correlated with the signal may be used in a joint measurement \textbf{B.}
	QTR: the target is in one among $m$ longitudinal space slots of size $\Delta x$. The returning signal arrives in a certain corresponding temporal time bin, while the other bins only detect the background. An ancillary-assisted strategy, exploiting a bipartite state and post-processing correlation measurements, may be used, represented in fainted colors. \textbf{C.} Uncorrelated noise  model: Each receivers intercepts a large number $M\gg1$ of low-mean-photon-number background modes. In this limit the noise and the probe photon statistics are approximated as independent.
	}\label{fig:Mod}
\end{figure*}

For the detection problem  schematically represented in Fig. \ref{fig:Mod}\textbf{A}, Tan \textit{et al.} \cite{Tan_2008} showed that 6 dB of quantum advantage can be obtained in the exponent of error probability decay with the probes usage's number \textit{L}, when compared to optimal classical case. The last is given by a coherent transmitter paired with homodyne detection.  
The most intriguing thing is that this advantage is obtained in strongly 'hostile' conditions: small mean signal photons' number $\mu\ll1$, large mean background $\mu_{B}\gg1$ and low target reflectance $\kappa\ll1$, a regime where typically delicate quantum states are not expected to be useful. 
However, after a large number of theoretical \cite{Nair_2011, Nair_2020, Yang_2022, Zhuang_2017a,Zhuang_2017c, Nair_2020} and experimental studies \cite{Lopaeva2013, Zhang_2013, Zhang_2015,Barzanjeh_2020} on different versions of QI, it appears unable to provide a convincing advantage in practical technological scenarios. First of all, the only known 6-dB-gain quantum receiver is exceedingly complicated and technologically out of reach \cite{Zhuang_2017d}, while feasible receivers based on OPA are limited to 3-dB advantage in the best case scenario \cite{Guha_2009}. Second, the decoherence-free storage of the ancillary mode in Fig. \ref{fig:Mod}\textbf{A} requires yet unavailable quantum memories. Finally, the aforementioned hostile regime in which QI finds the advantage requires very high time-bandwidth product of the transmitter to achieve meaningful detection probability. 

Surprisingly, only recently the problem QTR has been addressed \cite{Zhuang_2017c}. This problem can be tackled in different related formalism: as an estimation problem, establishing the ultimate range-delay accuracy in a continuous time measurement approach \cite{Zhuang_2022b}; In the framework of hypothesis-testing, namely as the task of determining which of many range-delay resolution bins contains a target that is known to be present in one of them \cite{Zhuang_2020a, Zhuang_2021}, as depicted in Fig. \ref{fig:Mod}\textbf{B}. Those analyses show a quantum enhancement of QTR similar to the QI's one in the same range of the parameters, leaving similar concerns on the practical utility.

Thus, at least for the optical case it worth reconsidering the QTR from a different perspective. Many real world LiDARs for 2D and 3D applications operate measuring the time of flight just by direct intensity measurement of short time puses \cite{Tachella_2019}. This is because homodyne detection requires phase-locking of the traveling signal with a local oscillator, which is usually hard in presence of environmental phase-noise, sample roughness and without information on the object distance. Phase-insensitive measurement are more practical and efficient, especially nowadays that single photon detector technology has few ps jitter \cite{Shin_2016}, corresponding to sub-millimeter resolution in space \cite{Morimoto_2020}. Thus, in many cases, the  optical classical receiver benchmark become to restrictive, rather it make sense to compare ranging schemes limited to phase-insensitive measurements , that means using direct photon counting (PC) receiver \cite{Lee_2023, Yang_2021}. Along the same spirit, the usual QI theoretical model for the noise, that combines coherently an highly populated single mode thermal state with the transmitted probe mode in a beam splitter-like model, is more suited for microwave regime rather than for the optical domain where blackbody thermal radiation is negligible at operating temperature and other sources of noise, like sun irradiance, with very low coherence time (of the order of fs) contribute. Thus, here we consider the QTR problem with a different model of noise background as represented in Fig. \ref{fig:Mod}\textbf{C}, where the detector integrates over a large number of independent optical background modes and the signal couples with just one of them, randomly. The mean photon number for each background mode is usually small, so that it be considered in the limit of Poisson photon statistics. 

The QTD with PC measurement has been previously consider in literature \cite{Liu_2019, Zhao_2022,Gregory_2019,England_2019}, however while strong time correlation in the photon entangled pairs is used to reduce the background effect, the classical transmitter is mostly taken without any narrow-temporal synchronization with the detectors, inhibiting any possibility of time filtering of the noise through time-gating. Referring to Fig. \ref{fig:Mod}\textbf{B}, in this work we consider the time resolution $\Delta t$ as fixed, for example set by the detector time bin (or jitter), and both quantum correlation coherence time and classical single shot signal pulse duration shorter than $\Delta t$. This ensure a fair comparison. Our analysts can be expended to schemes involving multiplexed frequency detection to further reduce the noise contribution \cite{Frick_2020,Blakey_2022}.

 %\section{Results}
   
\paragraph*{The Model.}

The schemes of target detection is shown in Fig. \ref{fig:Mod}\textbf{A}. A transmitter probe state $\rho_{T}$ is sent to detect a given partially reflecting target. A measurement is performed at the receiver on the returning signal state $\rho_{R}$. The target is immersed in a background in the state $\rho_B$ with mean photon number $\mu_B$. The effect of the signal interaction with the target and the background modes is described by a noisy and lossy channel $\mathcal{E}_{\kappa,\mu_{B}}$,  where $\kappa$ is the reflectance of the target. The resulting state at the receiver is $\rho_\kappa=\mathcal{E}_{\kappa,n_{B}}(\rho_T)$. In particular if  the target is absent, i.e. $\kappa=0$, only the background reaches the detector, so the final state is $\rho_{\kappa=0}=\rho_B$. Target detection is formally described as a binary hypothesis testing problem with hypotheses $\mathcal{H}_1:\rho_{R}=\rho_{\kappa}$ and $\mathcal{H}_0:\rho_{R}=\rho_B$, where hereinafter $\rho_{\kappa}$ is intended with $\kappa\neq0$.

On the other hand, in target ranging, the measurement are performed on $m$ time slots as depicted in Fig. \ref{fig:Mod}\textbf{B}. Since the modes in each time slot are considered independent, denoting as $\mathcal{H}_j$ the hypothesis of the target in the $j$-th time slot, the ranging becomes a $m-$ary hypotheses testing problem among $\{\mathcal{H}_1,...,\mathcal{H}_m\}$ defined through the corresponding output states:

\begin{equation}\label{m-ary H}
\mathcal{H}_j: \rho_{R}=\rho_{j}:=\rho_{\kappa}^{(j)}\bigotimes_{i\neq j}^{m-1}  \rho_B^{(i)}.
\end{equation}

where  $\rho^{(i)}$ is the state of the $i$-th time slot.

This formulation for the ranging problem assumes that the target is present in one and only one bin. Note that this assumption can be easily relaxed to various degree by adding more hypotheses to the set.
\vspace{0.5cm}

\paragraph*{QTR: Relation with QTD}

Clearly, detection and ranging are different tasks, nevertheless we can derive a connection between them, which holds  independently from the particular input states $\rho_{T}$ and form the interaction channel $\mathcal{E}_{\kappa,\mu_{B}}$ that can be either the conventional single mode beam-slitter-like mixing with a thermal noise contemplated in literature or our multimode optical noise channel of Fig.\ref{fig:Mod}\textbf{C}.

Given a binary hypothesis problem $\mathcal{H}_0$/$\mathcal{H}_1$ characterized respectively by the corresponding final states $\rho_0$/$\rho_1$ occuring with the corresponding prior probabilities $\pi_0 + \pi_1 =1$ the probability of error is given by the Helstrom bound \cite{Helstrom_1976}. 

%The Helstrom bound is guaranteed to be saturated by at least one POVM that can be theoretically constructed.

Consider now a sensing performed with a multi-copy input state  $\rho_{T}^{\otimes L}$, where the transmission channel $\mathcal{E}_{\kappa,\mu_{B}}$ is assumed to act independently on each copy. In this multi-copy case the Helstrom bound can be difficult do compute. However, in the asymptotic regime of  $L\to \infty$, an useful upper limit to the optimal probability of error is given by the Quantum Chernoff Bound: 
 \begin{equation}\label{Chernof bound}
 p_{err} (\rho^{\otimes L}_0,\rho^{\otimes L}_1) \leq \frac{1}{2} e^{-\xi_{QCB}L}
 \end{equation}
%\begin{equation}
%p_H (\rho_0,\rho_1)=\frac{1}{2}(1 - ||\pi_1 \rho_1 - \pi_0 \rho_0||)
%\end{equation}

where we fixed $\pi_i=1/2$, meaning that no prior information is avaliable.  The bound in Eq. (\ref{Chernof bound}) is tight in the error exponent for $L\gg1$ \cite{Audenaert_2007}. Here, $\xi_{QCB}$ is the \emph{Quantum Chernoff information} \cite{Audenaert_2007} defined as:

\begin{equation}
\xi_{QCB}(\mathcal{H}_0,\mathcal{H}_1):=\max_{\alpha \in [0,1]} C_\alpha(\rho_0,\rho_1) \label{eq:QCI}
\end{equation} 
i.e. by the maximization of the $\alpha$-information, $C_\alpha$, between the single-copy states $\rho_0$ and $\rho_1$: $C_\alpha(\rho_0,\rho_1):=-\log(\text{Tr}[\rho_0^\alpha \rho_1^{1-\alpha}]) \label{eq:a-C}$. $C_\alpha$ is concave in $\alpha$ over the interval $\alpha \in [0,1]$.  Notably, in the asymptotic limit $L\gg1$, the exponential decay rate for a set of $m$ hypotheses $\{\mathcal{H}_1,...,\mathcal{H}_m\}$, $\xi^{m}$, is equal to \citep{Li_2016}:
\begin{equation}
\xi^{(m)}_{QCB}=\min_{i,j}\xi^{(2)}_{QCB}(\mathcal{H}_i,\mathcal{H}_j) \label{eq:MCher}
\end{equation} 
In other words in the asymptotic regime, the exponential decay rate of multi-hypotheses testing is the same as the binary one of the two 'closest' hypotheses in the set. For the symmetry of the states in Eq. (\ref{m-ary H}), we can arbitrary chose two hypotheses of the set to substitute in Eq. (\ref{eq:MCher}), obtaining by Eq. (\ref{eq:QCI}):
\begin{align}
\xi_{TR}&= \max_{\alpha \in [0,1]} C_\alpha\left(\rho_{\kappa}\otimes \rho_B\bigotimes^{m-2} \rho_B , \rho_B\otimes \rho_{\kappa} \bigotimes^{m-2} \rho_B\right)\nonumber\\
&=\max_{\alpha \in [0,1]} C_\alpha\left(\rho_{\kappa}\otimes \rho_B, \rho_B\otimes \rho_{\kappa} \right)\nonumber\\
&=2\, C_{1/2}(\rho_{\kappa},\rho_B) :=2 \mathcal{B}_{TD}\geq  C_{\alpha}(\rho_{\kappa},\rho_B) \label{eq:tr}
\end{align}
where we introduced the quantum Bhattacharya information for target detection $\mathcal{B}_{TD}$. The last equality is demonstrated in the Supplemental material. Note that $C_{\alpha}$ is symmetrical at $\alpha=1/2$, i.e. $ C_{1/2}(\rho_{\kappa},\rho_B)= C_{1/2}(\rho_B,\rho_{\kappa})$.  The last inequality derives directly from the  concavity of the $\alpha$-information.  
Eq. (\ref{eq:tr}) gives a direct relationship between the optimal asymtotic performance in TR, given by the quantum Chernoff information $\xi_{TR}$ and the Bhattacharya information for target detection $\mathcal{B}^{\text{cla}}_{TD}$. The quantum Chernoff information for target detection writes according to Eq. (\ref{eq:QCI}) as $\xi_{TD}:=\max_{\alpha \in [0,1]} C_\alpha(\rho_B,\rho_{\kappa})$  one gets immediately that $\xi_{TR}\geq\xi_{TD}$, always. In other words Eq. (\ref{eq:tr}) states that asymptotically the task of ranging, given previous knowledge on the presence of the target,  can be performed with better accuracy than target detection, regardless of the number of time slots $m$ considered.
This result is completely general, meaning that it does not depend on the form of the probe states and on the kind of measurement performed.

\vspace{0.5cm}

\paragraph*{QTR: Phase-Insensitive Measurement.}

\begin{figure}
	\includegraphics[width=\columnwidth]{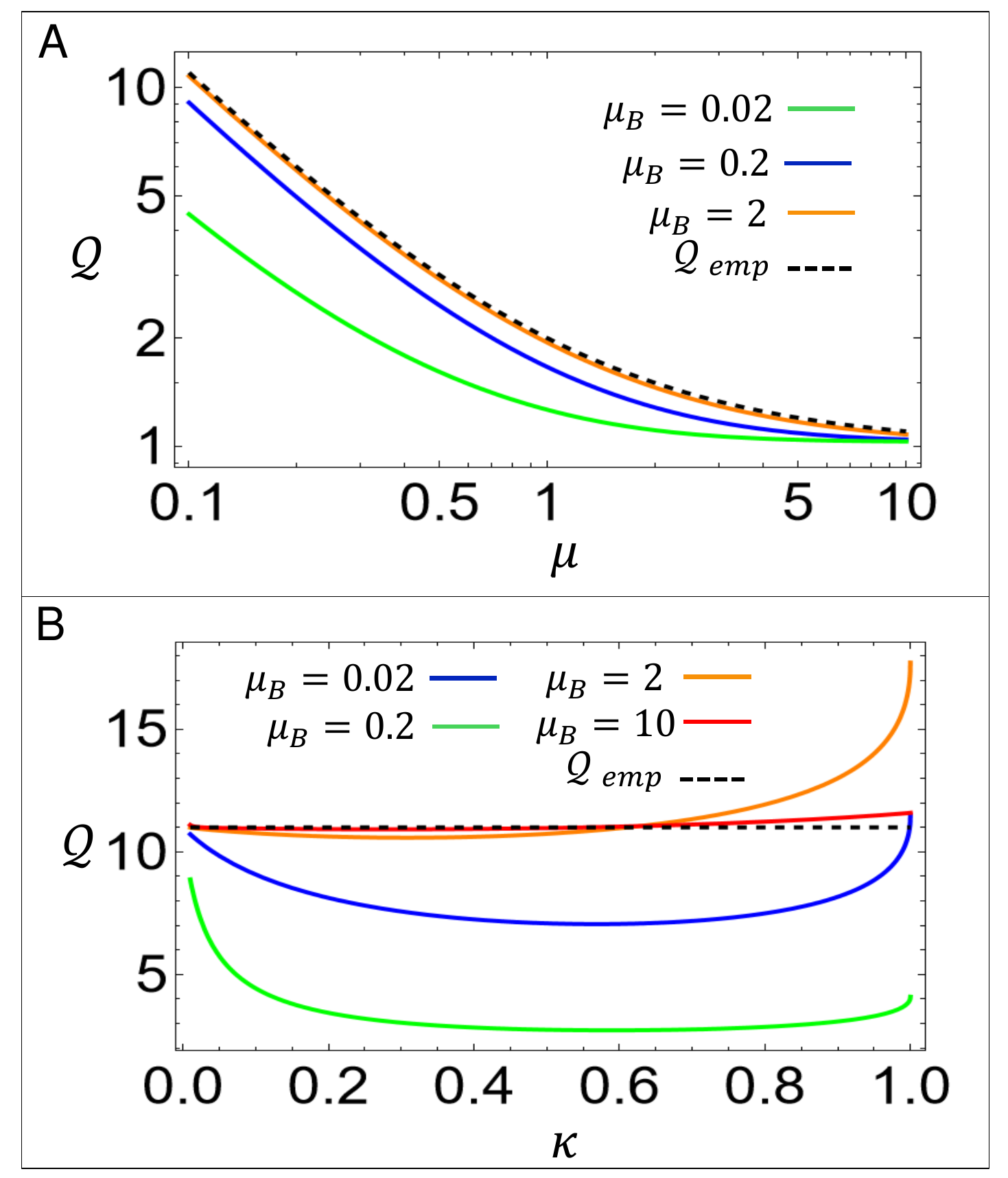}
	\caption{\textit{Asymptotic quantum advantage.} \textbf{A.} The quantum advantage $\mathcal{Q}$ defined in the text is plotted against the mean number of probes' photons $\mu$ (target reflectance $\kappa=0.1$) . \textbf{B.} Quantum advantage as a function of the target reflectance $\kappa$. ($\mu=0.1$).  Colors represents different value of the background mean photon number $\mu_{B}$. The black-dashed line is the empirical limit defined in the text.}\label{fig:asy}
\end{figure}
As motivated in the introduction, now we focus on the research of the quantum advantage when considering phase insensitive PC measurement, for both classical and quantum states of the transmitter. In particular we will compare the performance of a coherent transmitter $\rho_{T}=\rho_{coh}$, acting as a classical benchmark, with the non-classically correlated bipartite state $\rho_{T}=\rho_Q$ defined in the following. The QTR is performed using a multi-copy transmitter's state of the form $\rho_{T}^{\otimes L}$. Initially, let us consider $\rho_{T}$ a single mode state addressed to the target. For the target in the $j$-th time slot, the single-copy state at the receiver  $\rho_{j}$ is given in Eq. (\ref{m-ary H}). Given $|n_i\rangle$ the eigenstate with eigenvalue $n_i$ of the number operator $\hat{n}_i=\hat{a}_i^{\dagger} \hat{a}_i$ of the field in the $i$-th slot and defining the multimode Fock state $|\textbf{n}\rangle= \bigotimes_i |n_i\rangle$, the photo-counts probability distribution is then, $P_j(\mathbf{n})=\text{Tr}(\rho_j |\mathbf{n} \rangle \langle \mathbf{n}|)$. The best ranging performance will depend on the set of classical probability distributions $\{P_1(\mathbf{n}),..., P_m(\mathbf{n})\}$, corresponding to a classical $m$-hypotheses test on the set of measurement outcomes. The probability of error decays exponentially with $L$ at the rate fixed by the so called \textit{classical Chernoff information} $\xi^{(m)}_{CB}$. Similarly to the  measurement-independent (quantum) case of Eq. (\ref{eq:MCher}), the $m$-ary problem can be reduced to a binary test, where \cite{Leang_1997}:
\begin{equation}
\xi^{(m)}_{CB}=\min_{i,j}\xi_{CB}(P_i(\mathbf{n}),P_j(\mathbf{n})) \label{eq:mCherCla}.
\end{equation} 
The binary classical Chernoff information is defined as \cite{Nielsen_2013,Nielsen_2022}:
\begin{equation}
	\xi_{CB}(P_i,P_j):=\max_{\alpha \in [0,1]} -\log\bigg(\sum_{\mathbf{n}}P_i(\mathbf{n})^\alpha P_j(\mathbf{n})^{1-\alpha}\bigg) \label{eq:CCI}.
\end{equation}

Following the notation used for the states, let us label $P_B$ and $P_\kappa$ the photon number distributions at a given time slot, with only background or background plus returning signal respectively. Following similar steps used to derive Eq. (\ref{eq:tr}), starting form Eq. ($\ref{eq:CCI}$), one gets for the PC quantum target ranging:
\begin{equation}
\xi_{TR}^{\text{cla}}:=\max_{\alpha \in [0,1]} C^{\text{cla}}_\alpha (P_B P_\kappa, P_\kappa P_B)=2 \mathcal{B}^{\text{cla}}_{TD} \label{eq:CCBI}
\end{equation}

The detailed derivation of this result is given in the Supplementary Material. Here we have denoted quantities with the superscript "cla" to indicate that they refer to Chernoff, Bhattacharya and $\alpha$ information of probability distributions after fixing the measurement to PC, i.e. the classical version of the previously defined quantities for quantum states. In the following we will omit the superscript for brevity of notation. In our model for the noise, we always assume that the  background can be approximated having a Poisson distribution, $P_B(n)=\mathcal{P}_{\mu_{B}}(n)$, and the statistics of the photo-counts in the only slot receiving the probe's photons is the convolution of the background noise with the signal, i.e they sum up incoherently. For a the coherent state transmitter, $\rho_{T}=\rho_{coh}$, the probe's distribution is also a Poissonian $\mathcal{P}_{\mu}$,  so that $P_{k}(n)=\mathcal{P}_{\mu_{B}}\ast\mathcal{P}_{\mu}$ ('$\ast$' stand for convolution). In this case, the Chernoff information of the target ranging task, can be found analitically, as reported in the Supplementary Material, as:
\begin{equation}
\xi_{coh}=\mu+ 2\mu_B-2\sqrt{\mu_B} \sqrt{\mu_B+\mu}
\end{equation}
where $\mu$ is the mean number of photons in the probe state $\rho_T$ and $\mu_B$ is the mean number of background photons.

The quantum probe that we will consider is a collection of a large number of Two mode squeezed vacuum (TMSV) states, $\rho_Q=(|TMSV\rangle\langle TMSV |)^{\otimes R}$. The TMSV state is defined as $|TMSV\rangle=\sum_n c^{\mu_0}_n |n,n\rangle_{S,I}\langle n,n|$, with $|c_n^{\mu_0}|^2$ a thermal distribution with mean $\mu_0$. This state is entangled and presents perfect correlation between signal and idler photon numbers. Taking $R$ copies of this state preserves the photon number correlation and in the limit of small $\mu_0$ and large $R$ gives a marginal photon-number distribution that is Poissonian instead of the thermal one of the single copy state \cite{Ortolano_2021}, improving the performance. For a fair comparison with the coherent case we set $R\mu_0=\mu$ so that both probes have the same number of signal photons. The results found above are easily extended to a bipartite system considering the joint distributions of signal and idler in place of the signal one, and a result analogous to Eq.(\ref{eq:CCBI}) holds, meaning that the Chernoff information $\xi_{{Q}}$, for the quantum probe $\rho_Q$, can be computed in terms of the Bhattacharya information of the detection task, and no optimization over $\alpha$ is needed. The evaluation of $\xi_{{Q}}$ is performed numerically.

Let us now introduce the quantum advantage $\mathcal{Q}$ defined as:
\begin{equation}
\mathcal{Q}:=\frac{\xi_{{Q}}}{\xi_{{coh}}}
\end{equation}
In Fig. \ref{fig:asy}\textbf{A} we report $\mathcal{Q}$ in the regime of small reflectance ($\kappa=0.1$) as a function of the mean number of probe photons $\mu$ for different values of background photons $\mu_B=0.02, 0.2,2$. Interestingly, it turns out that when $\kappa\ll1$ and $\mu_B\gg1$ (in practice $\mu_B=2$ seems enough to fulfill the condition) the quantum advantage reaches the empirical bound $Q_{emp}= 1+1/\mu$, represented by the black dashed line. It is interesting that $Q_{emp}$ corresponds to the square of the ratio among the maximum phase-sensitive field correlation, $\langle a_{1} a_{2}\rangle$, allowed for the  quantum, $\sim\sqrt{u(u+1})$, and classical, $\sim u$, bipartite Gaussian states, a quantity which is known to be the origin of the quantum advantage in Gaussian state QI \cite{Shapiro_2020}. However, note that in our scheme we do not use the classical benchmark of correlated classical state, rather a coherent state. Fig. \ref{fig:asy}.\textbf{A} show that in general the quantum advantage is significant in the region of $\mu\ll1$, eventually approaching $\mathcal{Q}=1$, i.e. no advantage, in the opposite regime. Fig. \ref{fig:asy}.\textbf{B} shows that the dependence of $\mathcal{Q}$ from the target reflectance $\kappa$ has a non trivial behavior. Only for $\mu=10\gg1$ (the red curve) the Q dependence from $\kappa$ is negligible. In the other situation,  the highest advantage is found in the two extreme regions of either very high or very low reflectance. Low-reflectance regime $\kappa\ll1$ is the one providing advantage in the conventional model of Quantum Illumination, while the high-reflectance regime,  $\kappa\gg1$ is where the quantum reading protocol \cite{Pirandola_2011,Ortolano_2021,Ortolano_2022,Ortolano_2021b} finds the best quantum gain.   
\vspace{0.5cm}

\paragraph*{QTD: Time slot dependence.}

The asymptotic behavior of the probability of error gives no information about the dependence on the number $m$ of time slots. This dependence can be derived by setting an optimal post-processing strategy to compute the exact probability of error.

\begin{figure}
	\includegraphics[height=0.7\columnwidth,width=\columnwidth]{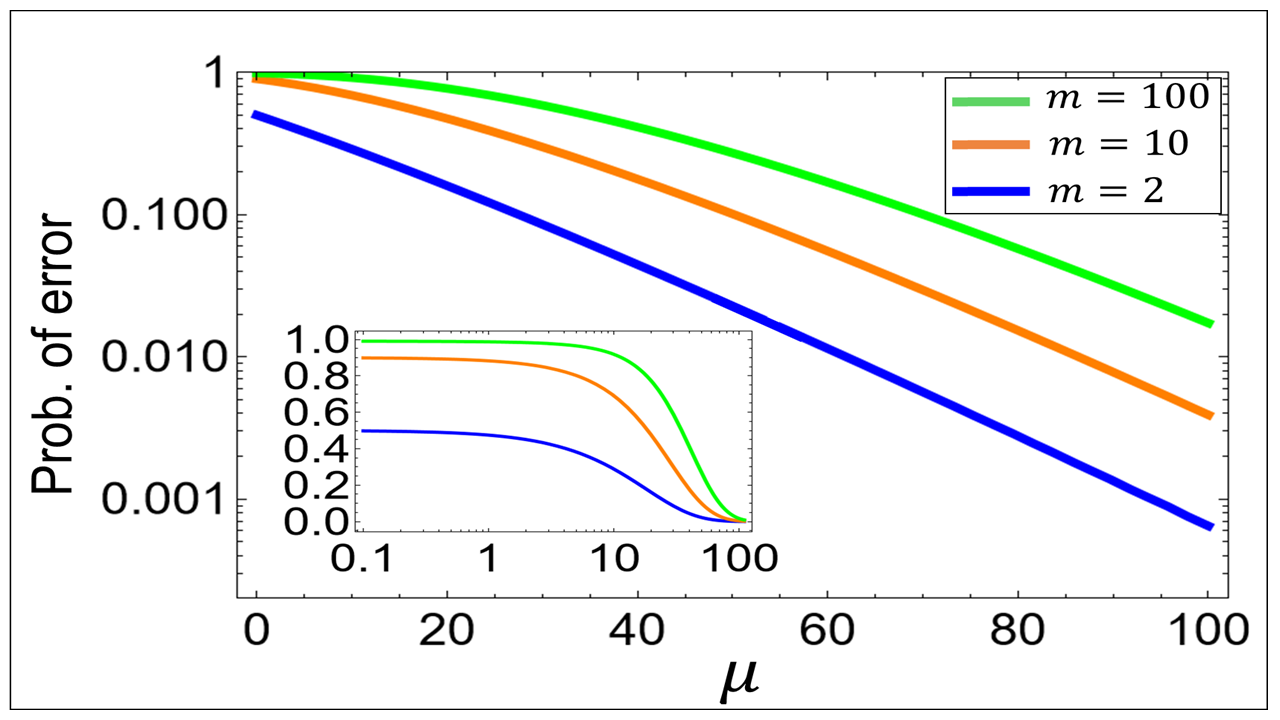}
	\caption{\textit{Error probability dependence from the number of time slots.} The probability of error  $p_{C}^{(1)}$ is reported as a function of the mean number of transmitted photons $\mu$. The three colors refer to different numbers $m$ of time slots according to the legend (target reflectance $\kappa=0.1$, mean number of background photons $\mu_B=1$). The main graph in linear-log scale, while the insets shows the same functions in log-linear scale. }\label{fig:mdep}
\end{figure}

Let us consider a single-shot scenario, i.e. let us fix $L=1$. For a coherent state transmitter the single-shot measurement outcome at the receiver is an array of photon counts $\mathbf{\tilde{n}}=\{\tilde{n}_1,...\tilde{n}_m \}$ referring to each of the $m$ time slots. Given $\mathbf{\tilde{n}}$ the best Baesyan decision is to select the time slot with the highest photon count \cite{Helstrom_1976}, selecting at random if the highest count is repeated in more than one slot. This yelds the following probability of error (see Supplementary Material):
\begin{equation}
p_{C}^{(L=1)}(\mu)=1-m^{-1}e^{-\kappa \mu}  G_m\left(1+\kappa \frac{\mu}{\mu_B}\right) \label{perrM}
\end{equation}
where we introduced the generating function $G_m(x)$:
\begin{equation}
G_m(x):=\sum_{n=0}^{\infty} x^n  \left(Q_{\mu_B}(n+1)^m-Q_{\mu_B}(n)^m\right) \nonumber
\end{equation}
$Q_{\mu}(n)=\Gamma(n,\mu)/\Gamma(n)$ being the regularized incomplete gamma function. 

In Fig. (\ref{fig:mdep}) we plot $p_{C}^{(L=1)}(\mu)$ for different numbers of time slots, $m=2,10,100$.  In the main graph, we show the curves in liner-log scale to highlight their similar asymptotic exponential decay coefficient in the limit of $\mu\gg1$, which results to be independent from $m$. Note that, as discussed later, in the particular case of coherent sates Eq. (\ref{asymt_coh}) holds,  thus the asymptotic slope is in fact determined by the Chernoff information.  

The insets of Fig. (\ref{fig:mdep})  shows, in a different scale,  how each curve starts with a probability of error of $1-1/m$ at $\mu=0$, corresponding to a random guess, as expected.

\vspace{0.5 cm}

\paragraph*{QTD: Non-asymptotic quantum advantage.}
The non-classically correlated probe $\rho_Q$ introduced in the asymptotic analysis of $L\gg1$ is unable to outperform the coherent state transmitter in ranging in the single-copy scenario with photon counting. This is due to the fact that knowledge of the idler does not change the optimal Bayesian decision, that remains to pick the higher signal count regardless of the idler. In other words, in the single-shot ranging with photon-counting, the performance depends only on the marginal signal photon number statistic. Thus a quantum advantage in this scenario can be obtained by a probe with sub-Poisson marginal distribution, such as the one in a Fock state. 

In this section we will consider instead a finite number $L$ of repetitions and explore the advantage of $\rho_Q$ in this non-asymptotic regime, that is the one of arguably most practical interest.
The outcome of the PC measurement in QTR scheme of Fig. \ref{fig:Mod}\textbf{B}, with the input state $\rho_Q^{\otimes L}$, is an array $\mathbf{\tilde{n}}_S=\{\mathbf{\tilde{n}}_1,...,\mathbf{\tilde{n}}_m\}$ for the signal, with each $\mathbf{\tilde{n}}_j$ being $L-$dimensional vectors of integers. For the idler the outcome is single $L-$dimensional vector $\mathbf{\tilde{n}}_I=\{\tilde{n}_I^{(1)},...,\tilde{n}_I^{(L)}\}$ with $\tilde{n}_I^{(i)}$ scalar integers.
In the following we show that an asymptotically optimal decision rule is to consider for each of the $m$ slots the scalar product $c_j:=\mathbf{\tilde{n}}_j \cdot \mathbf{\tilde{n}}_I$ and select as an outcome the slot $\tilde{j}=\arg\max{c_j}$.  Similar strategy for PC based QDR is known to be optimal \cite{Lee_2023}.  The probability of error of this stategy can be evaluated with numerical analysis and we denote it as $p_Q^{(L)}$.

On the other hand, for a coherent probe in the state $\rho_C$ the best stategy for $L-$copy state is to select the time slot with the highest total count after summing on the $L$ copies. In other words, for a probe consisting of $L$ copies of a coherent state each with $\mu$ mean photons achieves the same performance of a single coherent state with $L\mu$ mean photons, i.e. there is no advantage in spreading photons in different coherent modes and measuring them independently. It follows that the non-asymptotic multi-copy optimal probability of error, $p_C^{(L)}$, is calculated exploiting the result of Eq. (\ref{perrM}):
\begin{equation}\label{asymt_coh}
p_C^{(L)}(\mu)=p_C^{(1)}(L\mu)
\end{equation}

\begin{figure*}
	\includegraphics[height=0.8\columnwidth,width=\textwidth]{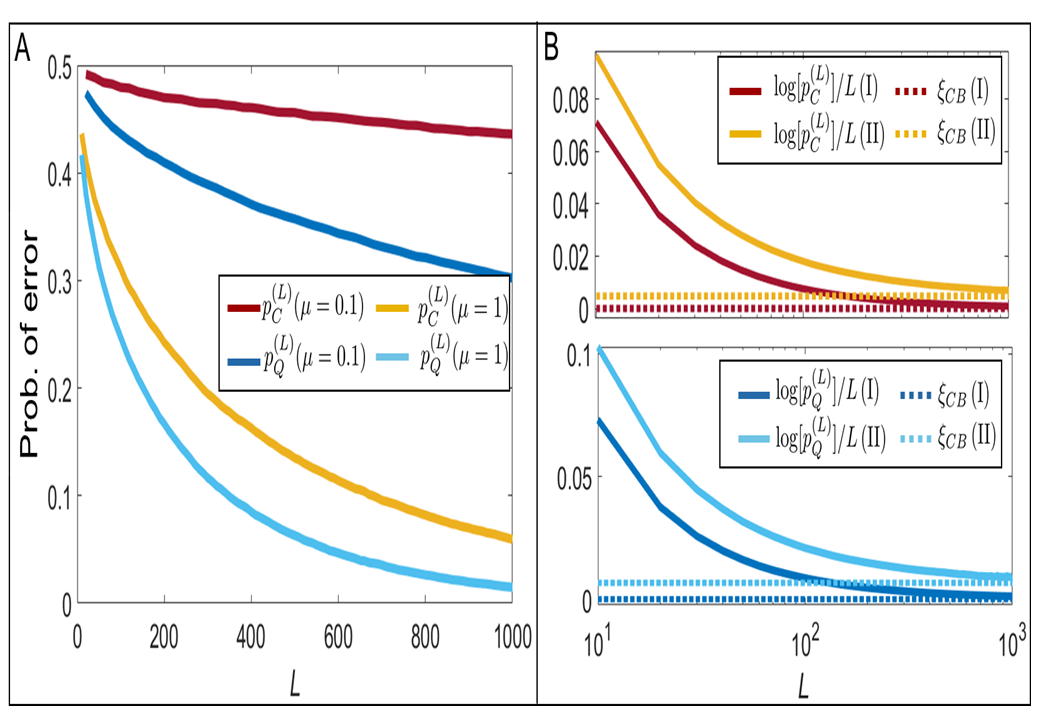}
	\caption{\textit{Non-Asymptotic probabilities of error.} \textbf{A.} Classical and quantum error probabilities, $p_C^{(L)}$ and $p_Q^{(L)}$ are compared in function of the number of probes' copies $L$ (here, $\mu_{B}=2,\kappa=0.1,\mu=0.1,1 $ according to the legend). \textbf{B.} Top panel shows the scaling of the logarithmic error probability $p_C^{(L)}$ (solid lines) with respect to corresponding the Chernoff bound (dotted lines), for two sets of parameters: (I) $\mu_{B}=2,\mu=2,\kappa=0.1$ and (II) $\mu_{B}=1,\mu=0.1 ,\kappa=0.1$.  The bottom panel shows the same kind of comparison this time of $p_Q^{(L)}$ and the corresponding Chernoff bounds with the same two sets of parameters, (I) and (II). }\label{fig:qadv}
\end{figure*}

In Fig. \ref{fig:qadv}\textbf{A} we compare the probabilities of error $p_C^{(L)}$ and $p_Q^{(L)}$ as a function of $L$, showing that a sensible quantum advantage can be found also in the non-asymptotic regime. Just for simplicity and without loss of generality, we refer to the case $m=2$. The comparison is shown for two values of the probe mean photon number ($\mu=0.1$ and $\mu=1$) while the other parameters are kept fixed. In Fig. \ref{fig:qadv}\textbf{B} we plot the logarithm of the error probabilities normalized to $L$ (solid lines), to show that they approach asymptotically the corresponding Chernoff informations $\xi_{{Q}}$ and $\xi_{{coh}}$ (dashed lines). Thus, being Chernoff bounds asymptotically tight, the corresponding practical decision strategies are asymptotically optimal. 

%\section{Discussion}
In this work we have shown that QTR is in general an easier task than QTD. Calculating the optimal bounds to the error probability in the asymptotic regime, for phase-insensitive measurement, we have found that QTR with an entangled transmitter can achieve an exponent advantage inversely proportional to the mean photon number of the probe with respect to the best classical transmitter if both are paired with PC receiver. Simple photon number correlation-like measurement is asymptotically optimal and its advantage persists also in the non asymptotic case. Our analysis represents a significant step in the analysis towards a real implementation of quantum LiDAR.    
\section*{Acknowledgments}
\subsection*{Funding}
This project has received funding from the European Defence Fund (EDF) under grant agreement 101103417 EDF-2021-DIS-RDIS-ADEQUADE.
Funded by the European Union. Views and opinions expressed are however those of the authors only and donot necessarily reect those of the European Union or the European Commission. Neither the European Union nor the granting authority can be held responsible for them.
\subsection*{Author contributions}
Both the Authors conceived the scheme of QTR paired with PC receiver. GO elaborated the theory and performed calculations and simulations. Both Author contributed to the paper writing. IRB is the responsible for funding.

\subsection*{Competing Interests}
The authors declare no competing interest.
\subsection*{Data availability} All data needed to evaluate the conclusions are reported in the paper. Further data are available under reasonable request to the corresponding author.

\bibliography{bib}
\onecolumngrid
\newpage
\setcounter{equation}{0}
\renewcommand\theequation{S\arabic{equation}}
\section*{Supplementary Material}
\subsection*{A. Chernoff information}
\paragraph*{Quantum Chernoff Information.}
To derive the quantum Chernoff Information of target ranging let us start from the definition in Eq.(3) of the main text. This definition, jointly with the symmetry of the hypotheses set $\{\mathcal{H}_m\}$ gives:
\begin{equation}
\xi_{TR}= \max_{\alpha \in [0,1]} C_\alpha (\rho_{\kappa}\otimes\rho_{B},\rho_B\otimes \rho_{\kappa}) \label{eq:cid} 
\end{equation}
Remembering that $C_\alpha(\rho_0,\rho_1)=-\log(\text{Tr}[\rho_0^\alpha \rho_1^{1-\alpha}])$, the $\alpha-$information in Eq.(5) can be expanded as:
\begin{align}
C_\alpha & (\rho_{\kappa}\otimes\rho_{B},\rho_B\otimes \rho_{\kappa})= \nonumber \\
&=-\log\left (\text{Tr}\left[\left (\rho_{\kappa}^{\alpha}\otimes\rho_{B}^{\alpha}\right)\left(\rho_B^{(1-\alpha)}\otimes \rho_{\kappa}^{(1-\alpha)}\right)\right]\right) = \nonumber \\
&=-\log\left (\text{Tr}\left[\left (\rho_{\kappa}^{\alpha}\rho_{B}^{(1-\alpha)}\right)\otimes\left(\rho_B^{(1-\alpha)} \rho_{\kappa}^{\alpha}\right)\right]\right)= \nonumber \\
&=-\log\left (\text{Tr}\left[\left (\rho_{\kappa}^{\alpha}\rho_{B}^{(1-\alpha)}\right)\right]\right)-\log\left (\text{Tr}\left[\left (\rho_{\kappa}^{(1-\alpha)}\rho_{B}^{\alpha}\right)\right]\right) = \nonumber \\
&=C_\alpha (\rho_{\kappa},\rho_{B}) + C_{1-\alpha} (\rho_{\kappa},\rho_{B}) \label{eq:ci1} 
\end{align}
where the last equality follows directly from the definition of $C_\alpha$. The concavity of $C_\alpha$ in $\alpha \in [0,1]$ means that:
\begin{equation}
C_{\omega \alpha + (1-\omega)(1-\alpha)}\geq \omega C_{\alpha}+ (1-\omega)C_{1-\alpha}
\end{equation} 
$\forall  \omega \in [0,1]$. Setting $\omega=1/2$ in the above concavity condition yields 
\begin{equation}
2 C_{1/2} \geq C_{\alpha}+ C_{1-\alpha} \label{eq:con}
\end{equation}
The maximization in Eq.(\ref{eq:cid}) is solved using Eq.(\ref{eq:ci1}-\ref{eq:con}) to yield:
\begin{equation}
\xi_{TR}= 2 C_{1/2} (\rho_{\kappa},\rho_{B}):=2\mathcal{B}_{TD}\label{eq:ci2}
\end{equation}
that is the result showed in the main text. Here we introduced the quantum Bhattacharya information for target detection, $\mathcal{B}_{TD}$, that is the $\alpha$-information at $\alpha=1/2$. Eq.(\ref{eq:ci2}) establishes an exact relation between the asymptotically optimal performance in ranging, expressed by $\xi_{TR}$, and the Bhattacharya information of the related task of target detection. Moreover given that the optimal performance in target detection is found by the maximization of $C_{\alpha} (\rho_{\kappa},\rho_{B})$, i.e. $\xi_{TD}:=\max_{\alpha \in [0,1]} C_{\alpha} (\rho_{\kappa},\rho_{B})$, Eq.(\ref{eq:con}), jointly with Eq.(\ref{eq:cid}-\ref{eq:ci1}), gives the inequality $\xi_{TR}\geq\xi_{TD}$. As discussed in the mean text this inequality shows that asymptotically the task of target ranging can be performed better than detection, assuming prior knowledge on the presence of the target.

\paragraph*{Classical Chernoff Information.}
The asymptotically optimal performance in distinguishing probability distributions (defining measurement outcomes), $P_0$ and $P_1$, is given by the classical Chernoff information, $\xi^{\text{cla}}$, defined in the binary case as:
\begin{align*}
\xi^{\text{cla}} (P_0,P_1)&:= \max_{\alpha \in [0,1]} C^{\text{cla}}_\alpha (P_0,P_1) \\
C^{\text{cla}}_\alpha (P_0,P_1):=& -\log\bigg(\sum_{\mathbf{n}}P_B(\mathbf{n})^\alpha P_\kappa(\mathbf{n})^{1-\alpha}\bigg)
\end{align*}
For multi-hypothesis testing the classical Chernoff information reduces to the lowest binary one \cite{Leang_1997}, similarly to the quantum case. In the case of target ranging the classical Chernoff information is given by:
\begin{align}
\xi_{TR}^{cla}:=\max_{\alpha \in [0,1]} C^{\text{cla}}_\alpha (P_B P_\kappa, P_\kappa P_B)
\end{align}
The independence of the time slots is expressed by product of independent distributions in place of the tensor product of states of the quantum scenario. With similar algebra to Eq.(\ref{eq:ci1}) we get, once again due to the concavity on $\alpha \in [0,1]$:
\begin{align}
\xi_{TR}^{\text{cla}}&=2  C_{1/2}^{\text{cla}} (P_{\kappa},P_{B}):= 2 \mathcal{B}^{\text{cla}}_{TD} \label{eq:claDR}\\
\xi_{TR}^{\text{cla}}&\geq \xi_{TD}^{\text{cla}} \nonumber
\end{align}
where we have introduced the classical Bhattacharya information for target detection, $\mathcal{B}^{\text{cla}}_{TD}$, also for the classical case. 
\paragraph*{Classical Chernoff Information for the coherent transmitter.}
We use Eq.(\ref{eq:claDR}) to find a closed form for the target ranging Chernoff information for a coherent state transmitter, signaling $\mu$ mean photons. As detailed in the main text the photon number distributions for measurement performed in slots with or without the target are both Poissonian. Specifically, if the target is present in the time slot the distribution is $\mathcal{P}_\kappa$, with mean $\kappa \mu+ \mu_B$, or  if only background photons are measured the distribution is $\mathcal{P}_B$, with mean $\mu_B$. 

The $\alpha$-information, $C_\alpha(\mathcal{P}_1,\mathcal{P}_2)$ for univariate Poisson distribution, with mean values $\mu_1$ and $\mu_2$, has the following closed form \cite{Nielsen_2013,Nielsen_2022}:
\begin{equation}
C_\alpha(\mathcal{P}_1,\mathcal{P}_2)=\mu_2+\alpha(\mu_1-\mu_2)-\mu_1^\alpha\mu_2^{1-\alpha} \label{eq:ainfo}
\end{equation}
The Chernoff information, denoted in the main text as $\xi_{coh}$, can be computed using Eq.(\ref{eq:claDR}) and Eq.(\ref{eq:ainfo}):
\begin{equation}
\xi_{coh}= 2 C_{1/2}(\mathcal{P}_\kappa,\mathcal{P}_B)=2\mu_B+\kappa\mu-2\sqrt{\mu_B}\sqrt{\mu_B+\kappa\mu}
\end{equation}
\subsection*{B. Non-Asymptotic probability of error}
The coherent transmitter analyzed in the main text sends $\mu$ photons, Poisson distributed, to perform target ranging. After the interaction with the target having reflectance $\kappa$ the photons arriving at the receiver will be still Poisson distributed but with a mean value scaled by $\kappa$, i.e. having mean value $\kappa \mu$. According to our model, at the receiver there is incoherent mixing of the returning signal with a collection of background modes, collectively Poisson distributed with mean $\mu_B$. Thus, each of the $m$ time slots receivers will either record photons from the signal mixed with background, Poisson distributed according to $\mathcal{P}_\kappa$, with mean $\kappa \mu+ \mu_B$, or background only photons, again Poisson distributed according to $\mathcal{P}_B$, with mean $\mu_B$. To perform target ranging after photon counting one needs to identify among the $m$ counts $\{n_1,..., n_m\}$ which one is most likely generated by the signal distribution $\mathcal{P}_\kappa$. In this single shot scenario the best strategy is to select the slot $i$ with the higher count, $i=\argmax{n_i}$, selecting at random if there is more than one slot with the highest count [Helstrom].
Assuming, without loss of generality that the signal is on the first slot, $i=1$, and using the independence of the time slots, the probability of success, $p_s$ of this strategy is given by [Helstrom]:
\begin{align*}
p_s=\mathbb{E}_{n_1}\Big[&\prod_{i=2}^m P(n_i<n_1) + 2^{-1} g_1 P(n_2=n_1)\prod_{i=3}^m P(n_i<n_1) + ... \\ 
&+m^{-1} P(n_2=...=n_{m-1}=n_1)g_{m-1} P(n_m<n_1)\Big]
\end{align*}
where $\mathbb{E}_{n_1}$ denotes the expectation value over the variable $n_1$ and $g_{k-1}$ accounts for the possible combinations and is defined as:
\begin{equation}
g_{l-1}= {{m-1}\choose {l-1}} 
\end{equation}
For our scenario the background only slots are identically distributed according to $\mathcal{P}_B$ so that:
\begin{equation}
P(n_i<n_1)=\frac{\Gamma(n_1,\mu_B) }{\Gamma(n_1)}
\end{equation}
and:
\begin{align*}
&p_s=\mathbb{E}_{n_1}\Bigg[\sum_{l=1}^m \frac{1}{l} {{m-1}\choose {l-1}}  \mathcal{P}_B(n_1)^l \left(\frac{\Gamma(n_1,\mu_B) }{\Gamma(n_1)} \right)^{m-l}\Bigg]= \\
&=m^{-1} \mathbb{E}_{n_1}\Bigg[ \mathcal{P}_B(n_1)^{-1} \sum_{l=1}^m  {{m}\choose {l}}  \mathcal{P}_B(n_1)^l \left(\frac{\Gamma(n_1,\mu_B) }{\Gamma(n_1)} \right)^{m-l}\Bigg]= \\
&= m^{-1} \mathbb{E}_{n_1}\Bigg[ \mathcal{P}_B(n_1)^{-1} \left(\left(\mathcal{P}_B(n_1)+\frac{\Gamma(n_1,\mu_B) }{\Gamma(n_1)}  \right)^m - \left(\frac{\Gamma(n_1,\mu_B) }{\Gamma(n_1)} \right)^{m}\right)\Bigg]= \\
&= m^{-1} \mathbb{E}_{n_1}\Bigg[\mathcal{P}_B(n_1)^{-1} \left( \left(\frac{\Gamma(n_1+1,\mu_B) }{\Gamma(n_1+1)}  \right)^m - \left(\frac{\Gamma(n_1,\mu_B) }{\Gamma(n_1)} \right)^{m}\right)\Bigg]
\end{align*}
Introducing the regularized Gamma function $Q(n,\mu):=\Gamma(n,\mu) /\Gamma(n)$ we can write the probability of success as:
\begin{align*}
p_s&= m^{-1} \mathbb{E}_{n}\Bigg[\mathcal{P}_B(n_1)^{-1} \left(Q(n+1,\mu_B)^m - Q(n,\mu_B)^{m}\right)\Bigg]
\end{align*}
And using the fact that $\mathcal{P}_\kappa(n)\mathcal{P}_B(n)^{-1}=e^{-\kappa\mu} \left(1+\kappa \mu/\mu_B\right)^n$ we have:
\begin{align*}
p_s&= m^{-1}e^{-\kappa\mu} \sum_{n=0}^{\infty} \left(1+\frac{\kappa\mu }{\mu_B}\right)^n \left(Q(n+1,\mu_B)^m - Q(n,\mu_B)^{m}\right)
\end{align*}
Finally defining the generating function $G_m(x)$:
\begin{equation}
G_m(x)=\sum_{n=0}^{\infty} x^n \left(Q(n+1,\mu_B)^m - Q(n,\mu_B)^{m}\right)
\end{equation}
we get the probability of error:
\begin{equation}
p_{err}=1-p_s=1-m^{-1}e^{-\kappa\mu}G_m\left(1+\frac{\kappa\mu }{\mu_B}\right)
\end{equation}
That is Eq.(12) of the main text.
\end{document}